# Unveiling the thermal transport mechanism in compressed plastic crystals assisted by deep potential

Yangjun Qin[1,2], Zhicheng Zong[1,2], Junwei Che[3], Tianhao Li[1,2], Haisheng Fang[1], Nuo Yang[2, †]

1) School of Energy and Power Engineering, Huazhong University of Science and Technology, Wuhan 430074, China

2) Department of Physics, National University of Defense Technology, Changsha 410073, China

3) Department of Applied Physics, College of Science, Xi'an University of Science and Technology, Xi'an, 710054, China

†Corresponding E-mail: N.Y. (nuo@nudt.edu.cn)

# Abstract

The unique properties of plastic crystals highlight their potential for use in solid-state refrigeration. However, their practical applications are limited by thermal hysteresis due to low thermal conductivity. In this study, the effect of compressive strain on the thermal transport properties of plastic crystal $[(CH_3)_4N][FeCl_4]$ was investigated using molecular dynamic simulation with a deep neural network potential. It is found that a 9% strain along [001] direction enhances thermal conductivity sixfold. The underlying mechanisms are analyzed through vibrational density of states, spectral energy densities, and mean square displacements. The enhancement in thermal conductivity is primarily due to increased group velocity and reduced phonon scattering, driven by volume compression within the 0-1 THz. These findings offer theoretical insights for the practical application of plastic crystals in thermal management systems.

**Introduction**

Global warming and carbon neutrality initiatives have driven the demand for advanced technologies in energy harvesting, utilization, and refrigeration[1]. Plastic crystals (PCs) have emerged as a novel and highly promising material for solid-state refrigeration[2], due to their ability to undergo significant entropy changes during phase transitions in response to variations in temperature and pressure.

PCs exhibit substantial isothermal entropy changes during the barocaloric effect, marked by significant entropy shifts as their transition between ordered and disordered phases. The barocaloric effect[2] offers distinct advantages over conventional refrigeration techniques[3-5], including reduced driving pressures [6] and enhanced fatigue resistance. For example, neopentyl glycol[7], a representative organic PC, demonstrates a phase transition enthalpy of up to 389 J/kg/K[2] under a driving pressure of 91 MPa. Alongside neopentyl glycol, other key compounds such as glycerin, pentaerythritol, and tris(hydroxymethyl)aminomethane[8-10] have also shown promise for barocaloric applications. The pronounced isothermal entropy change in PCs is primarily driven by the formation and dissociation of hydrogen bonds during the phase transition[11]. However, the disordered structure of PCs usually results in significant phonon scattering, which lowers thermal conductivity and increases thermal hysteresis[12, 13]. Since the magnitude of thermal hysteresis directly affects refrigeration efficiency, balancing optimal phase transition performance with the minimization of thermal hysteresis remains a critical challenge in advancing solid-state refrigeration technologies.

Generally, there are some methods to improve the thermal conductivity of PCs, including pressure modulation[11], electric field modulation[14], and doping modulation[15, 16]. Pressure modulation alters the molecular arrangement and interactions within the crystal, resulting in a more compact structure that reduce phonon scattering and improve thermal conductivity. For instance, Wang et al. demonstrated a threefold increase in the in-plane thermal conductivity of pentaerythritol under a pressure of 3.5 GPa [17]. while Li et al. observed a 10% enhancement in the thermal conductivity of neopentyl glycol at 0.2 GPa [11]. Electric field modulation has also shown promise; Deng et al. reported a 53% increase in thermal conductivity of vinylidene fluoride through electric field polarization[14], which improves system ordering, reduces phonon scattering, and increases group velocity. Doping modulation, which involves introducing alternative molecules or atoms to modify the composition and structure of plastic crystals, can influence phonon scattering and thermal conductivity[15]. However, this method can also affect the phase transition properties of the PCs. Notably, $[(CH_3)_4N][FeCl_4]$[18] is a typical ionic crystal with the properties of PCs and ferroelectric material with applications in various fields. Despite this, the effects of strain on ionic PCs remain relatively unexplored, underscoring the need for further investigation.

The deep potential (DP) [19-22] molecular dynamics (MD) is a powerful tool to study the physic properties and uncover the microscopic mechanism. It combines the accuracy of density functional theory (DFT) [23, 24] with the computational efficiency of classical MD. The reliability of this method has been validated across various systems, including ferroelectric, ionic [25], high-entropy alloy, and organic systems[23, 26-28]. Therefore, the application of deep neural network models can provide critical insights into the mechanisms underlying heat transport in these materials.

This work developed a deep potential for plastic crystal [$(CH_3)_4N$][$FeCl_4$] to explore the effect of compressive strain along different directions on the thermal transport properties using the DPMD[29, 30]. Firstly, the accuracy of the potential function was validated by comparing results from DFT and DP. Secondly, the thermal conductivity of PCs was evaluated under varying strains applied along three directions. Thirdly, the vibrational density of states (vDOS) and the normalized spectral energy density (SED) distributions were calculated to further characterize the system. The primary aim was to investigate phonon transport properties under different strain conditions. Finally, the mean square displacement (MSD) was calculated to assess strain-induced structural changes. This study provides critical insights into the regulation of thermal conductivity in PCs through strain engineering.

**Deep potential and method of simulation**

The DP of [$(CH_3)_4N$][$FeCl_4$], comprising four phases, was developed using deep potential generator (DP-GEN) [31, 32] in a closed-loop automated approach. The DP-GEN framework iteratively explores the configurational space through three key steps: training, exploration, and labeling. The initial dataset was generated by perturbing the four phases of the crystal and performing ab initio molecular dynamics (AIMD). Such calculations were performed based on the density functional theory (DFT) implemented in the Vienna Ab initio Simulation Package (VASP), in which the Perdew–Burke–Ernzerhof (PBE) generalized gradient approximation (GGA) and projector augmented wave (PAW) pseudopotentials were applied. As detailed in Fig. S2, the energy cutoff and k-point spacing were set to 520 eV and 0.4 respectively, ensuring accurate results.

After obtaining the initial data, four DP models were trained using different random number. More details of methodology can be referred to the supplement material. In the exploration step, the structural phase space was explored through MD simulation utilizing the Large-scale Atomic/Molecular Massively Parallel Simulator (LAMMPS)[33-35], which were conducted at various temperatures and pressures across different phase structures using the isothermal-isobaric (NPT) ensemble. In the labeling step, the remaining three potential functions were evaluated by comparing their predictions for energy, force, and virial properties. Structures with the highest force deviations were specifically selected for analysis:

$$\delta_f^{max} = max_i \sqrt{\langle |F_i - \langle F_i \rangle|^2 \rangle} \qquad (1)$$

When $\delta_{low} < \delta_f^{max} < \delta_{high}$, the configuration is labeled as a candidate configuration, and added to the initial dataset for training in the next iteration. The iterative process constitutes until the accuracy of the potential function reaches 99%. The distribution of the final maximum force deviation is shown in Fig. S1. Throughout the multiple iterations, approximately 0.4 million phase space structures were explored.

Once a high-accuracy potential function is established, the energy and force data of both the training and test were compared in Fig. 1. A strong linear correlation between the DP and DFT data is evident in Fig. 1(a) and Fig.1(b), confirming the high accuracy of the potential function. Additionally, the phase IV cell, containing 44 atoms, was used to calculate the dispersion curve using the phonopy[36] software to generate a $3 \times 3 \times 3$ supercell with 1188 atoms. Calculating this using the finite displacement method is exceedingly challenging and impractical. Therefore, the energy profile of the fourth phase was calculated independently and found to be in excellent agreement with DFT in Fig.1(c). This confirms the accuracy of potential function and its suitability for calculating the properties of plastic crystals.

The thermal conductivity κ were calculated using the equilibrium molecular dynamics (EMD) method based on the Green-Kubo formula [37]:

$$\kappa = \frac{V}{3k_B T^2} \int_0^\infty \langle J(0)J(t)\rangle dt \qquad (2)$$

where $\langle J(0)J(t)\rangle$ is the heat current autocorrelation function (HCACF), $k_B$ is Boltzmann constant, T is the temperature; V is the volume of the simulation cell. The heat current J was calculated using the following expression:

$$J = \frac{1}{V}\left[\sum_i e_i v_i + \frac{1}{2}\sum_{i<j}(F_{ij} \cdot (v_i + v_j))r_{ij}\right] \qquad (3)$$

where $e_i$, $v_i$, $r_{ij}$ and $F_{ij}$ are the energy, velocity, position vector and force, respectively.

The vDOS was calculated from the Fourier transform of the velocity autocorrelation function:

$$vDOS = \int_0^{\tau_0} \frac{v(0) \cdot v(t)}{v(0) \cdot v(0)} \exp(-2\pi i\omega t)dt \qquad (4)$$

where $\tau_0$ is the simulation time, $v$ is the velocity of atoms.

The formula for SED[38-41] determines the distribution of the vibrational energy in the wavevector–frequency space, calculated from the following expression:

$$\Phi(\mathbf{\kappa},\omega) = \frac{1}{4\pi\tau_0 N_T} \sum_{\alpha}\sum_{b}^{B} m_b \left| \int_0^{\tau_0} \sum_{n_{x,y,z}}^{N_T} u_\alpha\binom{n_{x,y,z}}{b};t) \times exp\left[i\mathbf{k}\cdot\mathbf{r}\binom{n_{x,y,z}}{0} - iwt\right]dt \right|^2 \quad (5)$$

Where $N_T$ represents the number of unit cells in PCs, $n_{x,y,z}$, B, $m_b$ and $u_\alpha$ are the number of the protocells, the total number of atoms in the protocells, the masses of the atoms, and the velocities in the direction α, with the velocities being taken to be x,y,z, respectively. And **r** is the vector of displacements from the origin to the protocells.

The MSD is calculated in terms of the following expression:

$$MSD(t) = \langle |r_i(t) - r_i(0)|^2 \rangle \quad (6)$$

where t is the simulation time, r represents the position of the atom.

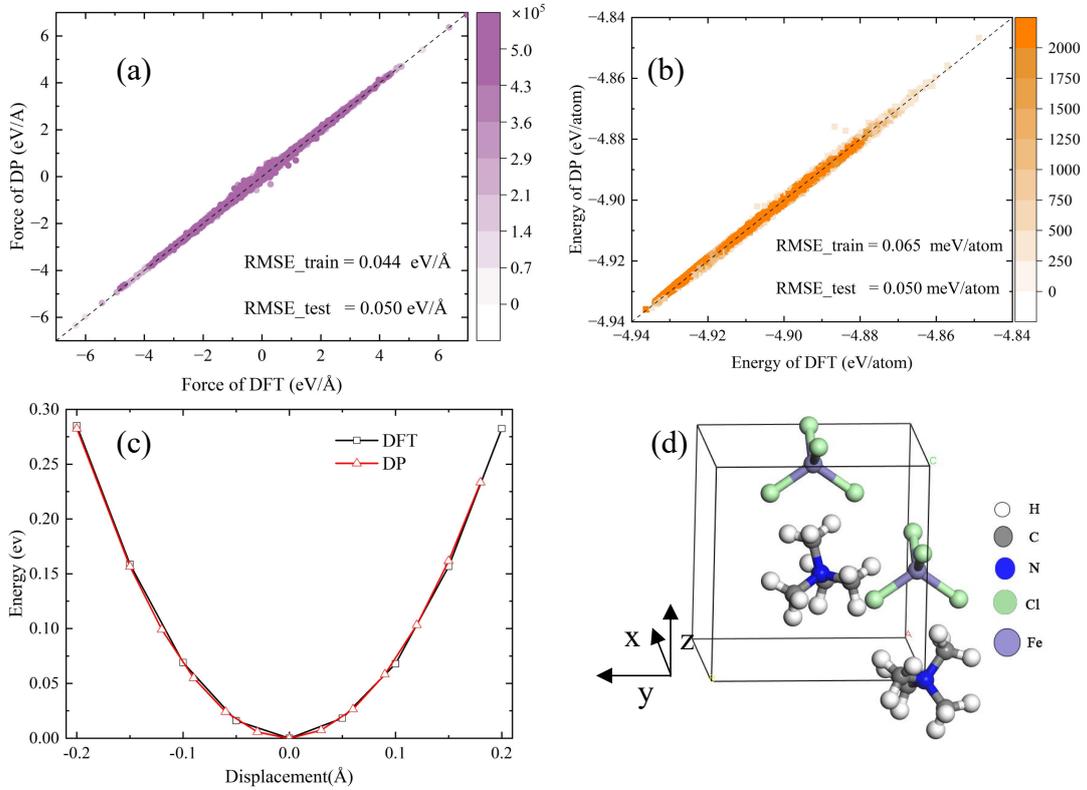

Figure 1. Training and validation results of deep potential. (a) Comparison of forces calculated

by the DP and DFT. (b) Comparison of energies calculated by the DP and DFT. (c) A comparison is presented of the energies obtained by DFT and DP, when Fe atom moves along the [100] direction. (d) Schematic structure of the system as visualized with OVITO[42].

**Results and discussion**

**The impact of strain along three different directions on the potential energy and volume of PCs was examined.** Modeling was performed using LAMMPS, with detailed relaxation and simulation parameters provided in the supplementary material. The changes in potential energy and volume at various strain levels were calculated, and the results are presented in Fig. 2(a) and Fig. 2(c). It can be seen that, as strain increases, the potential energy initially decreases with increasing strain, indicating enhanced stability of the system. However, as the strain continues to increase, the potential energy begins to rise, signaling reduced stability in the PCs. This trend persists until the potential energy reaches a critical peak, at which point the system collapses under the applied force. The structural configuration in the absence of strain is depicted in Fig. 2(b)-2(d). A comparison of the trends across the three directions highlights significant anisotropy in $[(CH_3)_4N][FeCl_4]$.

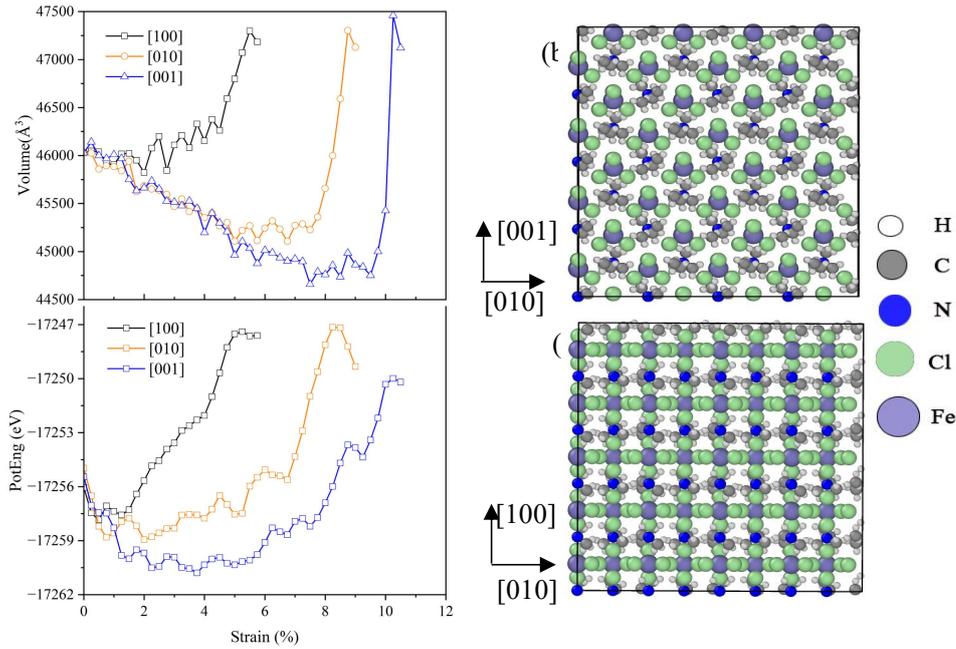

Figure 2. (a) and (c) are the volume and potential energy diagrams along different directions of strain, respectively, (b) and (d) depict the structural diagram at 0% strain for different plane, respectively.

**The effect of strain on the thermal conductivity of plastic crystals in different orientations was also investigated.** The thermal conductivity for various orientations and strain levels are presented in Fig. 3(a)-(c) and compared with available experimental data[43]. Thermal conductivity was calculated averaging the results of 10 calculations after the HCACF stabilized, as shown in Fig. 3(d). In the absence of strain, the average thermal conductivity of $[(CH_3)_4N][FeCl_4]$ is 0.23 W/mK. The thermal conductivity of $[(CH_3)_4N][FeCl_4]$ prepared by experimental hot pressing (unidirectional pressurization) in the in-plane direction and in the normal direction are 0.32 W/mK and 0.24 W/mK, respectively, with an overall average value of 0.19 W/mK, which is in close aligned with those obtained in the present study. The slight discrepancies can be attributed to the presence of grain boundaries, domain walls, and

other structural features in the experimental material, which increase phonon scattering and reduce the overall thermal conductivity.

The thermal conductivity of $[(CH_3)_4N][FeCl_4]$ was further evaluated under compressive strain along different orientations. Along the [100] direction, the effect of compressive strain on thermal conductivity was not significant, aligning with the observed changes in strain and volume. Interestingly, compressive strain applied along the [010] direction leads to a significant increase in thermal conductivity, with respective increases of 69%, 168%, and 8% along the [100], [010], and [001] directions, respectively. Notably, strain along the [001] direction results in an initial decrease in thermal conductivity, followed by a subsequent increase across all directions. The minimum point in this curve aligns with the system's lowest potential energy, indicating its maximum stability under strain. Notably, thermal conductivity increased by up to 110%, 580%, and 114% along the [100], [010], and [001] directions, respectively. This substantial rise in thermal conductivity can be attributed to two primary factors: the volume reduction and strengthened interatomic forces.

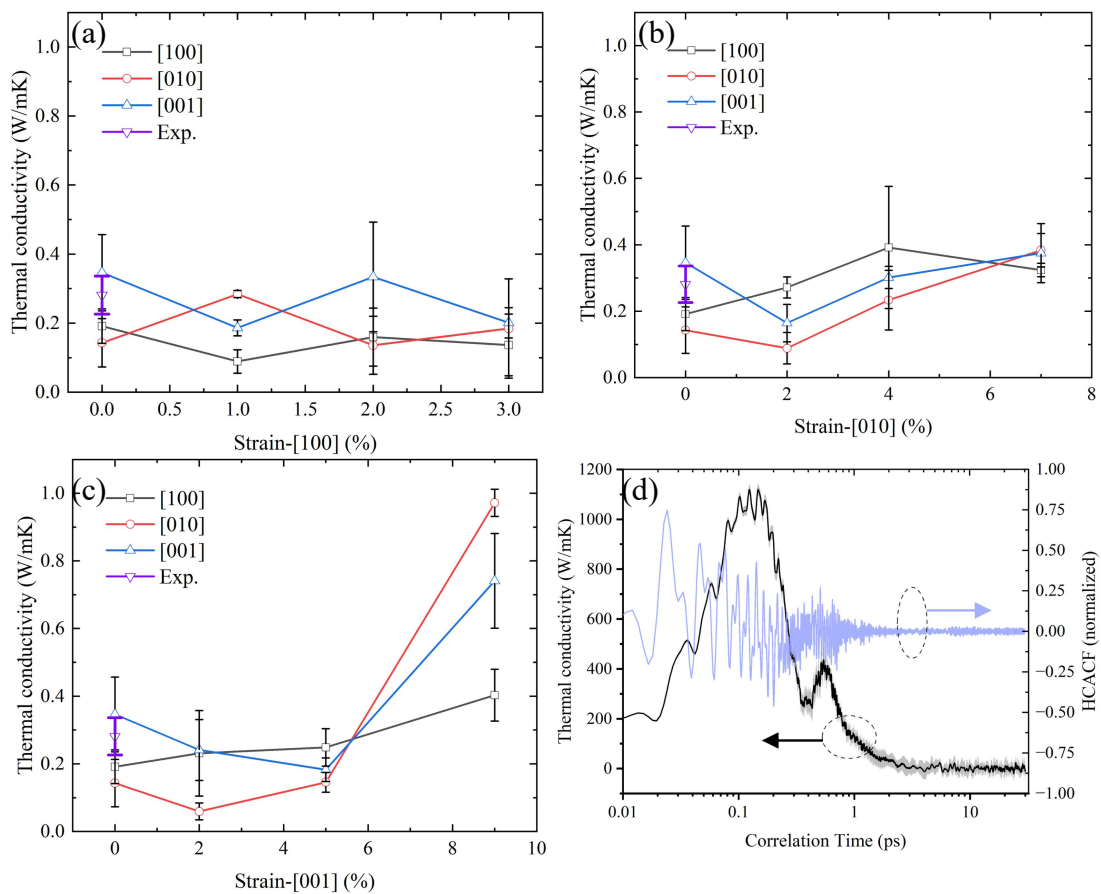

Figure 3. Thermal conductivity of PCs under the different strain. (a-c) The thermal conductivity of PCs along the [100],[010] and [001] direction under different strains. The violet triangle are experimental values for thermal conductivity averaged in three directions. (d) The normalized HCACF and the thermal conductivity of PCs as a function of correlation time.

To gain deeper insight into the mechanism by which strain affects thermal conductivity. the vDOS and normalized SED were calculated. Fig. 4 illustrates the distribution of phonon mode in the [(CH$_3$)$_4$N][FeCl$_4$] under various strain conditions. It can be seen that the most of phonon modes occur at frequencies below 50 THz, with additional modes near 90 THz attributed to the hydrogen atoms, whose lower mass leads to in higher vibrational frequencies. Interestingly, the number of phonon modes at different frequencies remains largely unchanged under strain, irrespective of direction, indicating that the system maintains its structural stability and integrity. In crystals, low-frequency phonon modes play a significant role in determining thermal conductivity. As shown in Fig. 4, in the low-frequency range, increasing strain leads to peak broadening, a blue shift in peak position, and an increase in group velocity—all of which enhance phonon transport. These changes collectively result in increased thermal conductivity within the system.

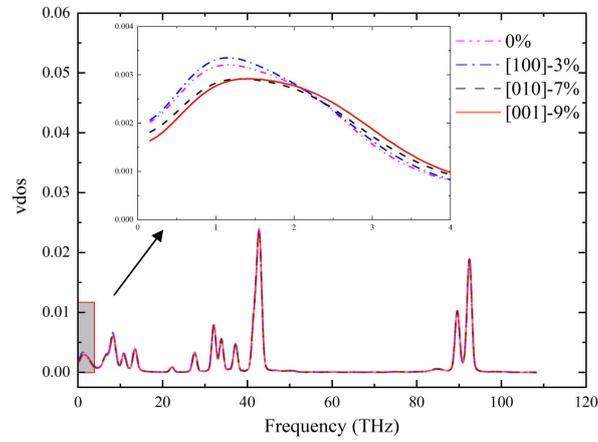

Figure 4. The vDOS of PCs with different strain

In addition to phonon modes, phonon scattering between individual modes is a key factor to determine thermal conductivity. To explore this effect, the SED distribution of $[(CH_3)_4N][FeCl_4]$ was calculated for strains within 0–1 THz frequency range. Fig. 5 presents the SED distribution at different strain levels, with frequency values displayed on the left and energy values on the right. Obviously, the system exhibits pronounced anharmonicity, which, combined with the presence of numerous protocell atoms, leads to significant phonon scattering, which accounts for the low thermal conductivity. Due to such a strong phonon scattering, distinct phonon modes are difficult to discern in Fig. 5. At strains of [010]-7% and [001]-9%, the SED shows increased vibrational energies in the low-frequency region, which corresponds to enhanced thermal conductivity. In contrast, at [010]-7% strain, the SED is more obscured, indicating increased phonon scattering, as seen in Fig. 5 and Fig. S2. This explains the reduction in thermal conductivity along the [010] direction compared to the [001] direction at 7% strain. The differences in phonon scattering and energy distribution at various strain levels also highlight the complex interplay between strain and phonon transport in determining thermal conductivity.

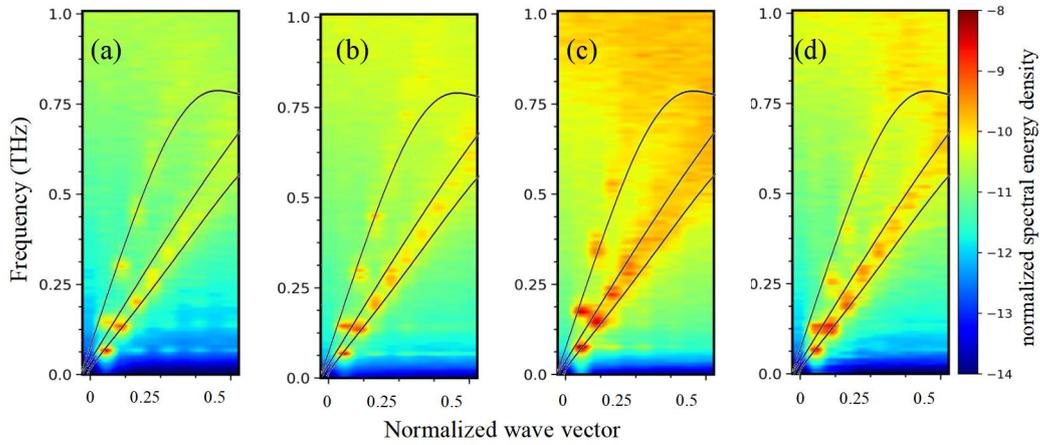

Figure 5. (a)-(d) The normalized SED of PCs with a strain of 0%, 3% ([100] direction), 7% ([010] direction) and 9% ([001] direction) at 0-1 THz. The black solid lines are the acoustic branches under 0% strain calculated by phonopy[36].

As shown in Fig. 6, the MSD demonstrates a tendency to fluctuate around a specific value in response to varying strain conditions, a behavior that is characteristic of crystals. The MSD distributions for different atoms in the absence of strain are depicted in Fig. 6(a), where it is evident that hydrogen atoms exhibit significantly higher MSD compared to other atoms. The heavier Fe atoms, along with the low-frequency vibrations of the N and Cl atoms, play a more prominent role in thermal transport. In Fig. 6(b), it illustrates the MSD variation in the [001] direction for Fe atoms under strain. An increase in the degree of compressive strain has been observed to result in a reduction in the MSD of Fe atoms. This trend shows an initial decline, followed by a slight recovery, with the MSD at 5% strain being the lowest and most stable within the system. The results show that strain consistently reduces MSD along the [001] direction, which contributes to the observed increase in thermal conductivity.

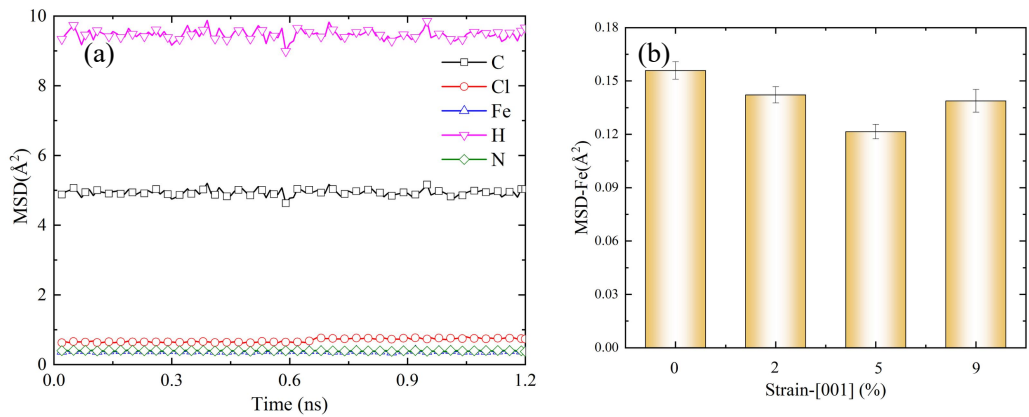

Figure 6. (a) MSD distribution of different atoms. (b) The MSD distribution of Fe atoms under different strain along [001] direction.

**Conclusion**

In a conclusion, a deep neural network potential model was developed to study the heat transport properties of $[(CH_3)_4N][FeCl_4]$ under varying strains. The deep potential was highly accurate, closely matching results from DFT calculations.

The thermal conductivity was computed by using MD simulations with the DP model. Notably, a 9% strain along [001] direction resulted in a six-fold increase in thermal conductivity along the [010] direction. To further investigate the mechanism by which strain impacts heat transport, the vDOS and SED were calculated. The findings revealed that high strain enhances low-frequency phonon modes, increases group velocity, and reduces phonon scattering.

Furthermore, the MSD under different strain conditions was calculated. The results revealed that the MSD of Fe, Cl, and N atoms is significantly lower than that of hydrogen atoms. The application of high strain mainly changes the vibrational modes of the massive atoms, weakens the phonon scattering, and enhances the thermal conductivity of the PCs.

These findings not only enhance the understanding of the heat transport in ferroelectric and plastic crystals under tensile stress but also offer guidance for future applications in thermal management and refrigeration systems.

**Conflicts of interest**

There are no conflicts of interest to declare

**Authorship contribution statement**:

Yangjun Qin: Investigation, Writing - original draft, Data curation, Formal analysis. Zhicheng Zong: Investigation, Software. Junwei Che: Investigation, Writing - original draft, Data curation. Tianhao Li: Investigation, Software. Haisheng Fang: Investigation, Nuo Yang: Project administration, Conceptualization, Writing - review & editing.

**Acknowledgements**

**Data availability**

The data that support the findings of this study are available from the corresponding author upon reasonable request.

# Supplemental Material

Unveiling the thermal transport mechanism in compressed plastic crystals assisted by deep potential

Yangjun Qin[1,2], Zhicheng Zong[1,2], Junwei Che[3], Tianhao Li[1,2], Haisheng Fang[1], Nuo Yang[2, †]

1) School of Energy and Power Engineering, Huazhong University of Science and Technology, Wuhan 430074, China

2) Department of Physics, National University of Defense Technology, Changsha 410073, China

3) Department of Applied Physics, College of Science, Xi'an University of Science and Technology, Xi'an, 710054, China

†Corresponding E-mail: N.Y. (nuo@nudt.edu.cn)

## Training details

For each of the four phases, 100 structure files were generated, with atomic positions perturbed by 0.001 nm. These structures were then subjected to ab initio molecular dynamics (AIMD) simulations at 300 K over 15 steps, yielding 4,500 initial data sets. The embedded network, fitted network, truncation radius, and rcut_smth were configured to (25, 50, 100), (240, 240, 240), 0.6 nm, and 0.5 nm, respectively. A total of 500,000 training steps were employed. The hyperparameters—start-pref_e, start-pref_f, start-pref_v, limit-pref_e, limit-pref_f, and limit-pref_v—were set to 0.02, 1000, 0.02, 1.0, 1.0, and 1.0, respectively.

Table S1: Exploration settings of DP-GEN iterations

| Iter. | Structures | Length(ps) | T(K) | Pressure (bar) | Candidate (%) | Accurate (%) | Failed (%) | Total data |
|---|---|---|---|---|---|---|---|---|
| 0 | PhaseIV(44 atoms)+phase III(44 atoms)+V（88 atoms）+VI(22atoms） | 0.1(200) | 200 300 330 360 420 | 0 10 100 1000 10000 | 31.3 | 65.40 | 3.31 | 21000 |
| 1 | | 0.1(200) | | | 9.49 | 88.64 | 1.87 | 21000 |
| 2 | | 0.1(200) | | | 3.45 | 96.55 | 0 | 21000 |
| 3 | | 0.1(200) | | | 3.55 | 96.45 | 0 | 21000 |
| 4 | | 0.1(200) | | | 1.64 | 98.36 | 0 | 21000 |
| 5 | | 0.25(500) | | | 43.89 | 56.00 | 0.10 | 51000 |
| 6 | | 0.25(500) | | | 15.05 | 84.78 | 0.16 | 51000 |
| 7 | | 0.25(500) | | | 12.15 | 87.81 | 0.03 | 51000 |
| 8 | | 0.5(1000) | | | 13.95 | 85.96 | 0.10 | 101000 |
| 9 | | 0.5(1000) | | | 12.59 | 87.30 | 0.12 | 101000 |
| 10 | | 0.5(1000) | | | 9.94 | 90.03 | 0.028 | 101000 |
| 11 | | 0.5(1000) | | | 4.86 | 95.12 | 0.02 | 101000 |
| 12 | | 2.5(5000) | | | 3.83 | 96.16 | 0.013 | 101000 |
| 13 | | 2.5(5000) | | | 1.44 | 97.65 | 0.91 | 101000 |
| 14 | | 7.5(15000) | | | 1.65 | 98.32 | 0.04 | 301000 |
| 15 | | 15(30000) | | | 1.48 | 98.52 | 0.0015 | 401000 |
| 16 | | 15(30000) | | | 0.65 | 99.35 | 7.4813 E-4 | 401000 |

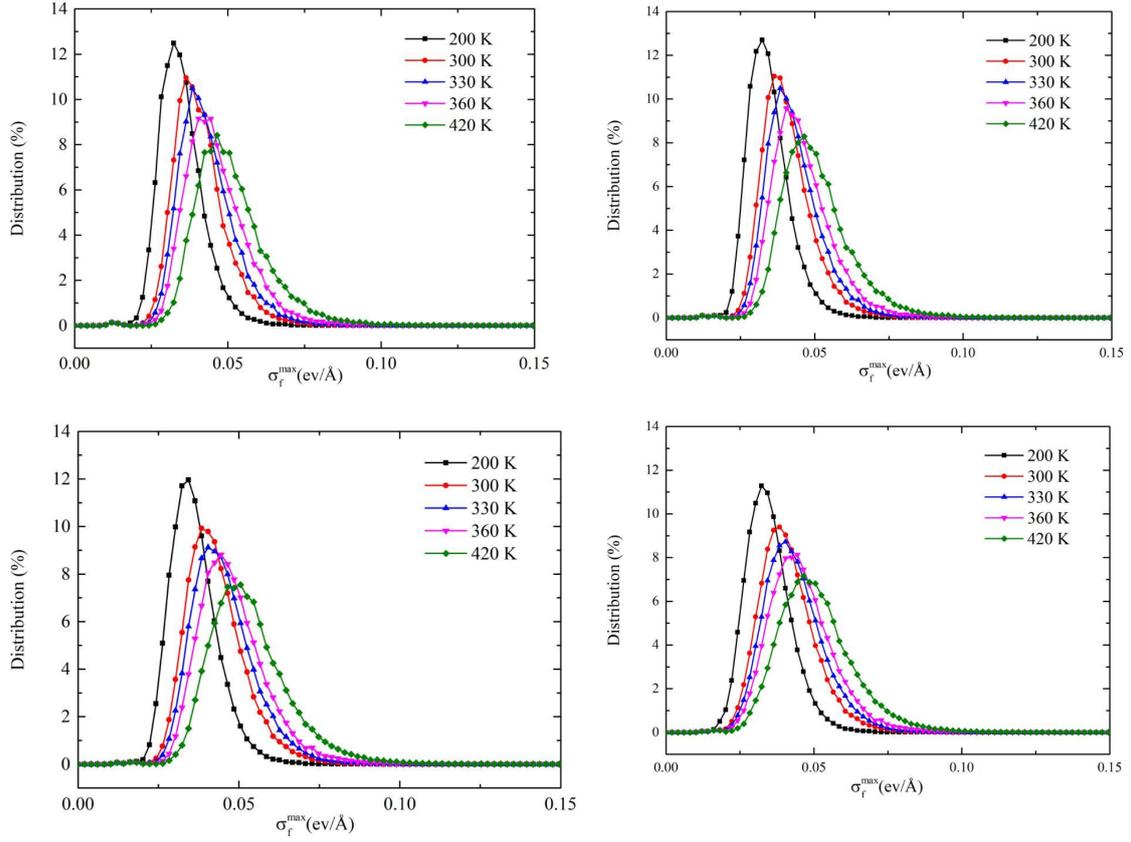

Figure S1 The maximum force deviation distribution for the four phases at the final iteration is presented herewith.

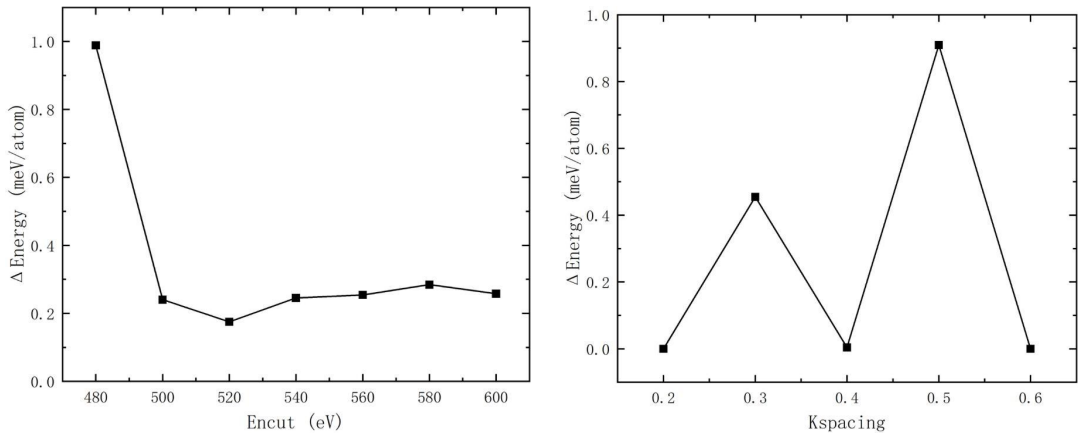

Figure S2 (a) The total energy differences to the kinetic energy cutoff; (b) The total energy differences to the k-point grid

**Simulation details**

All molecular dynamics simulations were conducted using the LAMMPS software. A 5*4*4 supercell was constructed with lengths of 3.55 nm, 3.58 nm, and 3.63 nm in the [100], [010], and [001] directions of the system, respectively, with periodic boundaries in all three directions. The velocity-Verlet method was employed to iteratively solve Newton's equations, thereby obtaining the position and velocity information of the atoms. The time step employed for the simulation is 0.5 fs. Prior to the application of the strain, the initial model is subjected to an energy minimization process. Subsequently, the NPT ensemble is relaxed at 330 K and 1 atm for a duration of 0.5 ns, thereby facilitating the complete release of stresses within the system. The system was subjected to uniform strains, resulting in strains of 3%, 7%, and 9% in the [100], [010], and [001] directions, respectively. Subsequently, a 300K relaxation is conducted under NVT ensemble for 0.3 ns, followed by an additional run under NVE ensemble for 0.2 ns. Finally, data collection and statistical analysis are performed to calculate the heat flow and thermal conductivity.

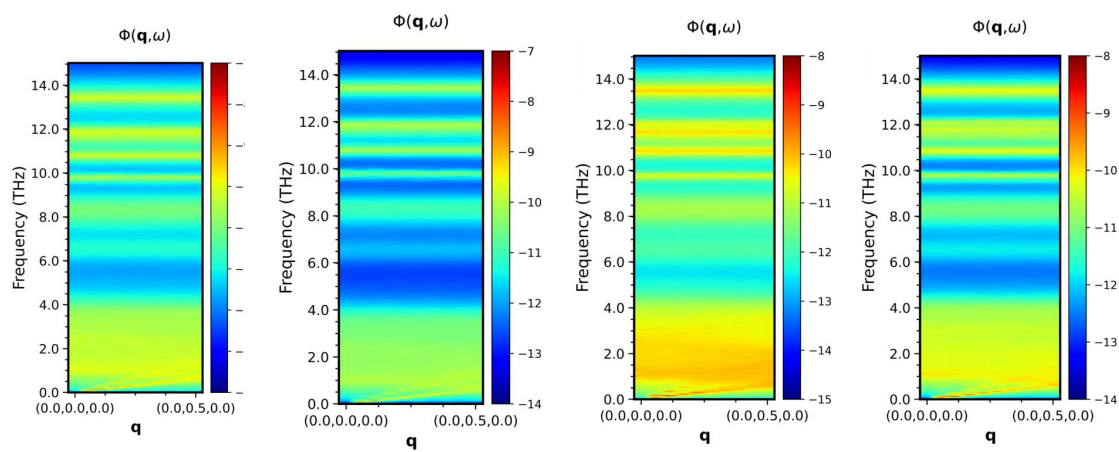

Figure S2. The normalized SED of PCs with strain of 0%, [100]-3%, [010]-7% and [001]-9% at 0-15 THz.